\documentclass[twocolumn,aps,floats,superscriptaddress]{revtex4}
\usepackage{graphicx}

\newcommand{\mr}[1]{{{\mathrm{#1}}}}

\newcommand{\fsd}{f_{\sigma}^{\dagger}}
\newcommand{\fs}{f_{\sigma}}
\newcommand{\s}{\sigma}
\newcommand{\w}{\omega}
\def\nun{\nu_n}
\def\wn{\omega_n}

\newcommand{\inte}{\int_0^\beta \!\!\!\! d\tau}

\begin{document}

\title{Exact scaling functions of the multichannel Kondo model}
\author{Serge Florens}
\affiliation{Institut f\"ur Theorie der Kondensierten Materie, Universit\"at Karlsruhe,
76128 Karlsruhe, Germany}

\begin{abstract}
\vspace{0.2cm}
We reinvestigate the large degeneracy solution of the multichannel Kondo
problem, and show how in the universal regime the complicated integral equations 
simplifying the problem can be mapped onto a {\it first order} differential
equation. This leads to an {\it explicit} expression for the full zero-temperature
scaling functions at - and away from - the intermediate Non Fermi Liquid fixed point, 
providing complete analytic information on the {\it universal} low - and intermediate - 
energy properties of the model. These results also apply to the widely-used Non
Crossing Approximation of the Anderson model, taken in the Kondo regime.
An extension of this formalism for studying finite temperature effects is
also proposed and offers a simple approach to solve models of strongly
correlated electrons with relevance to the physics of heavy fermion compounds.
\end{abstract}

\maketitle

Models of quantum impurities constitute a central piece in our understanding
of strongly correlated systems, as demonstrated by the importance of these ideas
for the physical description of magnetic impurities in metals \cite{hewson} or
superconductors \cite{Pan_imp_SC,PolkSachVoj}, the correlation-driven metal to
insulator transition \cite{RMP_DMFT}, the behavior of heavy fermion compounds 
\cite{revue_HF,schroeder}, and the coherent transport through
nanostructures \cite{revival_kondo}. Despite of their relative simplicity, one usually has
recourse to different techniques in order to fully grasp the complete physics
displayed by these localized hamiltonians. Unfortunately, analytical
approximation schemes devised to deal with the Kondo problem, like the 
slave boson method \cite{slave_boson2}, lead in general to too much simplifications.
On the contrary, purely numerical methods \cite{NRG_Anderson1} or exact 
techniques \cite{Bethe_multi} provide quantitative results, but often lack the flexibility to
solve more interesting cases, like lattice impurity models, or to compute dynamical 
observables. Conformal field theory (CFT) can also provide interesting insights on this
problem \cite{AffLud}, which is however limited to the elucidation of the critical behavior 
right at the possible fixed points.

Another way of tackling impurity hamiltonians lies in setting up controlled limits
that lead to a simplified solution encompassing all physical aspects of the problem. 
We will focus here on a large number of channels limit of quantum impurity models 
that was proposed previously \cite{OP_AG_multi2,OP_AG_multi1}, from
which the well-studied \cite{Cox_Ruck_multi,Bickers_RMP} self-consistent pertubation theory 
(Non Crossing Approximation or NCA) appears to be a limiting case. However
simplified, this alternative approach is known to lead to challenging non-linear 
integral equations, from which limited analytical information can be gathered 
\cite{OP_AG_multi2, mueller_hart}. A numerical solution of these equations is often possible, but
is increasingly difficult to obtain at low temperature and for real frequency
quantities.
Thus little work has been undertaken using this method
for the many interesting extensions of the Kondo model, like multi-impurity
hamiltonians or the Kondo lattice. This is unfortunate, as this technique 
offers a concrete opportunity for building a consistent theory for the competition of 
Kondo screening and magnetism \cite{OP_AG_multi1} which lies at the heart of the quantum 
criticality displayed in heavy fermions compounds.

We will show in this paper that much progress can be made towards both analytic
and numerical solutions of the multichannel Kondo model. Indeed we are able for
the first time to solve this problem at zero temperature in the universal regime 
(when the Fermi energy is large compared to frequency). The key step of this 
solution lies in the mapping of the saddle-point integral equations onto a much simpler
non-linear differential equation, that can be solved explicitely. Our calculation 
follows the well-known analysis by Mueller-Hartmann of the similar NCA equations for
the infinite $U$ Anderson impurity model \cite{mueller_hart}. 
However the Kondo regime of these equations seems not to have been considered previously 
in the literature, and constitutes the originality of the present work with respect 
to \cite{mueller_hart}.
We are thus able to obtain analytic expressions for the frequency-dependent 
scaling functions, for the Kondo temperature and for the precise location of the intermediate
Non Fermi Liquid fixed point. This offers also new insights on quantum criticality: we introduce
below a concept of complex Kondo temperature and describe an interesting crossover phase
diagram around the intermediate coupling fixed point. 
We will finally show how these ideas can be formulated at finite temperature, 
by deriving non-linear {\it finite difference} equations for the Green's functions, 
that may be useful to solve numerically more challenging extensions of the present 
problem.

Considering the Kondo model from now on, we will be interested here in a multichannel 
version that consists of a SU($N$) spin antiferromagnetically coupled 
to $K$ channels of conduction electrons, as expressed by the hamiltonian:
\begin{equation}
H = \sum_{k,\s,\alpha} \epsilon_k c^\dagger_{k\s\alpha} c_{k\s\alpha}
+ \frac{J_K}{N} \sum_{kk'\s\s'\alpha} S_{\s'\s} c^\dagger_{k\s\alpha} c_{k'\s'\alpha} 
\end{equation}
using a fermionic representation $S_{\s'\s} = f^\dagger_{\s'} f_{\s} - q_0
\delta_{\s\s'}$ of the localized spin, where $\s=1\ldots N$. The constraint
$\sum_{\s} \fsd \fs = q_0 N$ should also be enforced. Here the parameter $q_0 N$ 
may be understood as the size of the quantum spin, and the channel index $\alpha$ is
taken to run from $1$ to $K$.

In a previous work \cite{OP_AG_multi2}, Parcollet and Georges have shown how this model could be
simplified in the limit where both $N$ and $K$ are large, keeping $\gamma = K/N$
fixed (the fact that the spin size $q_0 N$ is taken to be large is also an
important point in obtaining their solution). Introducing the spinon propagator 
$G_f(\tau) = - \big< T_\tau \fs(\tau) \fsd(0) \big>$ and the Green's function 
$G_B(\tau) = + \big< T_\tau B_\alpha(\tau) B^\dagger_\alpha(0) \big>$ of the channel 
bosonic field $B_\alpha(\tau)$ conjugate to the operator $\sum_{k\s} c^\dagger_{k\s\alpha} f_{\s}$,
the saddle-point equations read (in imaginary time): 
\begin{eqnarray}
\label{eq:Sf}
\Sigma_f(\tau) &=& \gamma G_0(\tau) G_B(\tau) \\
\label{eq:SB}
\Sigma_B(\tau) &=& - G_0(-\tau) G_f(\tau) = G_0(\tau) G_f(\tau)
\end{eqnarray}
for a particle-hole symmetric bath Green's function 
$G_0(i\w) \equiv \sum_k 1/(i\w-\epsilon_k)$. The self-energies in the previous
set of equations are given by the Dyson equations:
\begin{eqnarray}
\label{eq:Gf}
G_f^{-1}(i\wn) &=& i\wn + \lambda - \Sigma_f(i\wn) \\
\label{eq:GB}
G_B^{-1}(i\nun) &=& \frac{1}{J_K} - \Sigma_B(i\nun)
\end{eqnarray}
where $\wn$ ($\nun$) denotes a fermionic (bosonic) Matsubara frequency. 
The Lagrange multiplier $\lambda$ is used to enforce the constraint related to the spin
size, which reads in the large $N$ limit:
\begin{equation}
G_f(\tau=0^-) = q_0
\end{equation}
The usual NCA equations \cite{Cox_Ruck_multi} are obtained when the constraint 
$\sum_{\s} \fsd \fs = 1$ is performed exactly, and this amounts to take the $q_0
\rightarrow 0$ limit in the saddle-point equations.
It is useful to analytically continue equations~(\ref{eq:Sf})-(\ref{eq:SB}), and
one readily obtains in terms of retarded quantities at zero temperature
\cite{SF_AG1}:
\begin{eqnarray}
\nonumber
\Sigma_f(\w) &=&  \gamma \int_0^{+\infty}  \frac{\mr{d}\epsilon}{\pi}
G_0''(\epsilon) \big[ G_B(\w\!-\!\epsilon) - J_K \big] \\
\label{eq:SfT0}
&& - \gamma \int_{\w}^{+\infty} 
\frac{\mr{d}\epsilon}{\pi} G_0(\epsilon) G_B''(\w\!-\!\epsilon) \\
\nonumber
\Sigma_B(\w) &=& \int_0^{+\infty} \frac{\mr{d}\epsilon}{\pi}
G_0''(\epsilon) G_f(\w-\epsilon) \\
\label{eq:SBT0}
&& - \int_{\w}^{+\infty}
\frac{\mr{d}\epsilon}{\pi} G_0(\epsilon) G_f''(\w-\epsilon)
\end{eqnarray}
where double primed quantities denote imaginary parts.
The subtracted $J_K$ in (\ref{eq:SfT0}) comes from a careful inspection of
the short time behavior of $G_B(\tau)$.
As we are interested in the universal regime where the Fermi energy of the
conduction electrons is the biggest energy scale, we follow \cite{mueller_hart} and
consider a flat density of states of half-width $\Lambda$: 
$G_0(\w) \simeq i G_0''(\w) = -i \pi \rho_0 \theta(\Lambda^2-\epsilon^2)$,
with $\rho_0 = 1/(2\Lambda)$ (the real part of $G_0$ can be discarded to order
$\Lambda^{-1}$). This leads in the previous equations to integrals running from 
the lower bound $\epsilon_\mr{min}=\w-\Lambda$ that can be replaced 
by $\epsilon_\mr{min} \simeq -\Lambda$ for frequency smaller than $\Lambda$. 
This trick allows to obtain simple expressions for the self-energies:
\begin{eqnarray}
\nonumber
\Sigma_f(\w) \! &=& \! - \gamma \rho_0 \int_{-\Lambda}^{\w} \!\!\! \mr{d}\epsilon 
\big[ G_B(\epsilon) - J_K \big]
+ i  \gamma \rho_0 \int_{-\Lambda}^{0} \!\!\! \mr{d}\epsilon \; G_B''(\epsilon) \\
\label{eq:SBfinal}
\Sigma_B(\w) \! &=& \! - \rho_0 \int_{-\Lambda}^{\w} \!\!\! \mr{d}\epsilon \; G_f(\epsilon)
+ i  \rho_0 \int_{-\Lambda}^{0} \!\!\! \mr{d}\epsilon \; G_f''(\epsilon)
\end{eqnarray}
that lead by differentiation to the following relations:
\begin{eqnarray}
\label{eq:dSf}
\frac{\partial}{\partial \w} \Sigma_f(\w) &=& - \gamma \rho_0 \big[G_B(\w) - J_K \big] \\
\label{eq:dSB}
\frac{\partial}{\partial \w} \Sigma_B(\w) &=& - \rho_0 G_f(\w)
\end{eqnarray}
Equation (\ref{eq:SBfinal}) imply also the following boundary condition
at the lower band edge (for $\Sigma_B$):
\begin{equation}
\label{eq:BSB}
\Sigma_B(-\Lambda) = i  \rho_0 \int_{-\Lambda}^{0} \!\!\! \mr{d}\epsilon \; G_f''(\epsilon)
\end{equation}

Following \cite{mueller_hart}, we now define the inverse propagators $Y_f(\w) \equiv - G_f^{-1}(\w)$ and 
$Y_B(\w) \equiv G_B^{-1}(\w)$. Combining the previous two relations with the
Dyson equations (\ref{eq:Gf})-(\ref{eq:GB}) for real frequencies, we easily arrive to:
\begin{eqnarray}
\frac{1}{Y_f} \frac{\partial}{\partial \w} Y_f &=& 
\frac{1-\gamma\rho_0 J_K}{\rho_0} \frac{\partial}{\partial \w} Y_B + \gamma \frac{1}{Y_B}
\frac{\partial}{\partial \w} Y_B \\
\label{eq:Yf}
Y_f^{-1} &=& - \frac{1}{\rho_0} \frac{\partial}{\partial \w} Y_B
\end{eqnarray}
The first equation provides a first constant of motion, that reads:
\begin{equation}
\label{eq:cst}
\log \big[-Y_f(\w)\big] + C = \frac{1-\gamma\rho_0 J_K}{\rho_0} Y_B(\w) + \gamma \log Y_B(\w)
\end{equation}
where $C$ is determined by the boundary condition (\ref{eq:BSB}).
The crucial difference between the multichannel Kondo model we consider in this
paper and previous work \cite{mueller_hart} on the NCA for the infinite $U$ Anderson model 
is that, by inserting relation (\ref{eq:Yf}) into (\ref{eq:cst}), we obtain a {\it first order} 
differential equation:
\begin{equation}
\label{eq:ED}
\frac{\partial}{\partial \w} Y_B = \rho_0 e^{C} (Y_B)^{-\gamma}
e^{(1-\gamma\rho_0 J_K) Y_B/\rho_0}
\end{equation}
instead of a non-linear second order differential equation. Ultimately, this
simplification is due to the disappearance of the high-energy scale associated
with charge fluctuations on the impurity. We emphasize however 
that this result also applies to the usual NCA of the Anderson model, 
in the {\it Kondo regime}.

To exhibit the exact solution of the multichannel Kondo model, we 
introduce two functions $f^{\pm}_\gamma(x)$ defined by the integral
(which is easily computed for integer $\gamma$):
\begin{equation}
x = \int_0^{f^\pm_\gamma(x)} \mr{d}t \;e^{\pm t} \; t^\gamma
\end{equation}
and thus obtain the following scaling form of the inverse bosonic
propagator:
\begin{eqnarray}
\label{eq:scale}
Y_B(\w) &=& A f^{a}_\gamma(\w/z_K) \\
\mr{where} \;\;\; a &=& \mr{Sign}(1-\gamma\rho_0 J_K)
\end{eqnarray}
The numerical prefactor in the previous equation is given by
$A \equiv \rho_0/|1-\gamma \rho_0 J_K|$, and a {\it complex} Kondo temperature
$z_K \equiv \rho_0^\gamma |1-\gamma \rho_0 J_K|^{-1-\gamma} e^{C} $ has
to be introduced. The fact that the propagators scale with respect to a complex
energy scale stems from the presence of a spectral asymmetry for generic values
of $q_0$ \cite{OP_AG_multi2}.

\begin{figure}[htbp]
\begin{center}
\includegraphics[width=7.5cm]{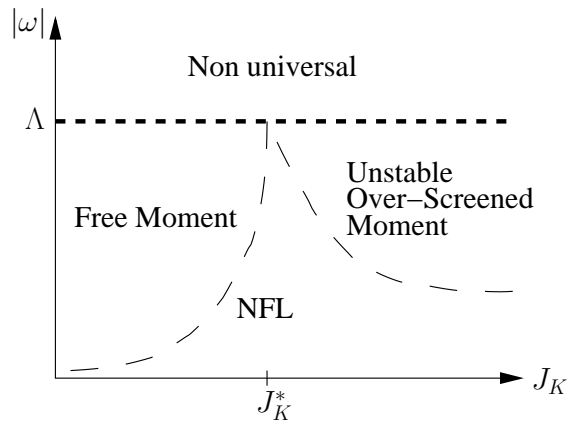} 
\end{center}
\caption{Zero temperature $(\w,J_K)$ crossover diagram: the dashed line
represents the Kondo temperature $T_K$ below which Non Fermi Liquid (NFL)
behavior is observed; above this line two {\it universal} crossover regimes
described in the text are characterized by the scaling functions~(\ref{eq:scale}). 
For $|\w| \gtrsim \Lambda$, non universal corrections start to appear.}
\label{fig:fixedpoint1}
\end{figure}

Equation (\ref{eq:scale}) can be considered to be the complete solution of the 
large number of channels Kondo model for all energy scales below the high energy cut-off, and
constitutes the central result of our paper. In particular, it is valid for
frequency below {\it and} above the the Kondo temperature, defined as $T_K
\equiv |z_K|$, as long as $T_K$ is much smaller than the cut-off $\Lambda$. 
To analyze the properties of this scaling function is first particularly
simple right at the Non Fermi Liquid fixed point. Indeed, equation~(\ref{eq:ED})
allows to locate the coupling $J_K^* = 1/(\rho_0 \gamma)$ at which the
various propagators exihibit scale invariance:
\begin{equation}
\label{eq:FP}
Y_B(\w) = \big[ \rho_0 (1+\gamma) e^{C} \w \big]^{1/(1+\gamma)}
\;\; \mr{at} \;\; J_K = J_K^*
\end{equation}
for all frequencies smaller than the cut-off. 
We note also that, away from the fixed point $J_K^*$, the behavior
(\ref{eq:FP}) still applies, albeit for $\w < T_K$.
The power law obtained in equation~(\ref{eq:FP}) agrees with previous work
\cite{mueller_hart,OP_AG_multi2}, in which the zero-temperature low-frequency propagator 
$G_B(\w) \propto \w^{2\Delta_B-1}$, with $2\Delta_B = \gamma/(1+\gamma)$, was
obtained.
For frequencies comparable to $T_K$ and away from the fixed point,
equation~(\ref{eq:scale}) leads to {\it universal} corrections to this 
pure power law behavior, a result that could not be obtained from the low frequency 
analysis performed in \cite{OP_AG_multi2}.
The various regimes observed are expressed graphically in figure~\ref{fig:fixedpoint1}.
To understand first the universal crossover arising at $J_K<J_K^*$,
we consider large frequency above $T_K$, and get $Y_B(\w) \propto \log |\w|$, 
so that $G_f(\w) \propto 1/\w$ from equation~(\ref{eq:Yf}). This corresponds to a 
free moment regime, where the impurity spin is weakly bound to the screening cloud 
of the conduction electrons. The scaling function $f_\gamma^+(x)$ thus describes 
the crossover from weak coupling to the fixed point $J_K^*$ \cite{Noz_Bland}.
From the same token, $f_\gamma^-(x)$ is associated to the flow from strong to
intermediate coupling; we note however that in the case $J_K>J_K^*$ one has
roughly $T_K \sim \Lambda/\Gamma(\gamma+1)$, so that the condition $T_K \ll
\Lambda$ (and hence universality of this regime) is met only for $\gamma \gg 1$.
This picture of the various crossover regimes that we have obtained is particularly
explicit, and illustrates the renormalization flow around the attractive fixed point.
To make contact with the theory of quantum critical phenomena \cite{sachdev_book}, 
we emphasize that $J_K$ is for the multichannel Kondo model an irrelevant perturbation 
around $J_K^*$, so that no phase transition occurs as $J_K$ is varied. 
Therefore the low energy behavior at all $J_K$ is controlled by
the intermediate fixed point, in contrast to the case of a relevant tuning parameter driving
a true quantum phase transition.

\begin{figure}[htbp]
\begin{center}
\includegraphics[width=7.5cm]{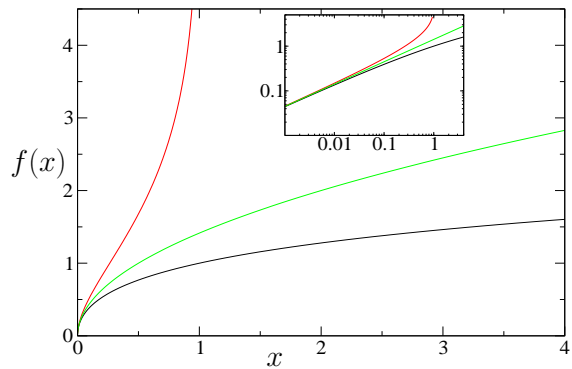} 
\end{center}
\caption{Plot on both linear and logarithmic scales of the scaling functions 
$f^+_\gamma(x)$ (lower curve) and $f^-_\gamma(x)$ (upper curve) for $\gamma=1$, 
as compared to their low energy asymptotic form $[(1+\gamma) x]^{1/(1+\gamma)}$ (middle
curve).}
\label{fig:plot_scaling2}
\end{figure}

Another way to depict the crossovers taking place around $J_K^*$ is to plot the 
scaling functions $f_\gamma^\pm(x)$, figure~\ref{fig:plot_scaling2}.

It is also possible to compute the exact value for the Kondo temperature 
$T_K$. We will do this in the limit $q_0 \rightarrow 0$, which corresponds 
actually to the usual Non Crossing Approximation.
In this case, the pseudo-particle propagators get maximally asymmetric, and the
spectral functions $G_B''(\w)$ and $G_f''(\w)$ vanish for all $\w<0$. This
implies an explicit expression for the boundary condition (\ref{eq:BSB}), $\Sigma_B(-\Lambda) = 0$,
and a purely real inverse propagator $Y_B(\w)$ (for negative frequency).
The initial condition $Y_B(-\Lambda) = 1/J_K$ gives for $J_K<J_K^*$ the
following relation:
\begin{equation}
\frac{T_K}{\Lambda} = \left[ \int_0^{1/(\rho_0 J_K) - \gamma} \; \mr{d}t \; 
e^t \; t^\gamma
\right]^{-1}
\end{equation}
In the limit of small $J_K$, one finds:
\begin{equation}
T_K \simeq \Lambda (\rho_0 J_K)^\gamma e^\gamma e^{-1/(\rho_0 J_K)}
\end{equation}
i.e. a vanishingly small Kondo temperature, as expected. When $J_K$ approaches
the fixed point value $J_K^*$, $T_K$ is maximum and Non Fermi Liquid properties
apply up to the high energy cutoff.

We now discuss temperature effects. In \cite{OP_AG_multi2} was shown 
how to obtain from the large-$N$ equations~(\ref{eq:Sf}-\ref{eq:GB}) the exact 
{\it finite-temperature} low-frequency (i.e. $\w \ll T_K$) scaling functions 
using CFT arguments. It seems unfortunately impossible to apply this approach 
to extend our exact scaling function~(\ref{eq:scale}), valid for 
$T_K \sim \w \ll \Lambda$, to finite temperature.
However we can succeed in generalizing the derivation of the linear differential
equations~(\ref{eq:dSf}-\ref{eq:dSB}) to the case of non zero temperature. For
that purpose we consider the {\it imaginary time} saddle-point equations in the
scaling regime:
\begin{eqnarray}
\nonumber
\Sigma_f(i\wn) &=& - \gamma \inte e^{i\wn \tau} 
\frac{\pi \rho_0}{\beta \sin(\pi\tau/\beta)} \big[G_B(\tau) -J_K \big]\\
\label{eq:SBw}
\Sigma_B(i\nun) &=& - \inte e^{i \nun \tau} 
\frac{\pi \rho_0}{\beta \sin(\pi\tau/\beta)} G_f(\tau) 
\end{eqnarray}
This gives the idea of writing a {\it finite difference} equation for the
self-energies, which provides indeed a remarkably simple result:
\begin{eqnarray}
\nonumber
\!\!\!\!\!\frac{\Sigma_f(i\nun+i\pi/\beta) - \Sigma_f(i\nun-i\pi/\beta)}{2\pi/\beta} 
\!\!\!&=&\!\!\!  - i \gamma \rho_0 \big[G_B(i\nun) -J_K \big]\\ 
\label{eq:SB_finite}
\!\!\!\!\!\frac{\Sigma_B(i\wn+i\pi/\beta) - \Sigma_B(i\wn-i\pi/\beta)}{2\pi/\beta} 
\!\!\!&=&\!\!\! -i \rho_0 G_f(i\wn) 
\end{eqnarray}
Although these relations do not lead to much analytical understanding as
compared to their zero temperature counterpart, they offer an economical way 
of tackling the finite temperature integral equations, which may be of interest
for some generalizations of the Kondo model. 

We would like to conclude the paper on the possible applications of the formalism 
discussed in this work. Our main result, the computation of the full zero-temperature
scaling functions of the many channel Kondo model, offers perspectives both in
the physics of quantum impurities and strongly correlated systems. First,
questions that we will address in future work concern the direct comparison of the
exact scaling function~(\ref{eq:scale}) to the numerical solution of the
large-N integral equations~(\ref{eq:Sf}-\ref{eq:GB}), using both the standard
numerical routines with fast Fourier fransforms and the 
system of finite difference equations~(\ref{eq:SB_finite}).
In the same direction, it would also be interesting to perform a similar
analysis and gain some analytical insight on the multichannel Kondo model 
in the bosonic representations \cite{OP_AG_multi1}, a model that allows a richer phase diagram 
with a transition from over-screening to underscreening.

As was pointed out in the introduction, models of several quantum impurities or
lattice versions of the Kondo hamiltonian are of great experimental
importance and notoriously difficult to tackle theoretically. We hope that the
analytical step undertaken in the present work will allow further progress in this
direction.

\begin{acknowledgements}
The author thanks A. Georges, A. Rosch and M. Vojta for useful comments on the
manuscript.
\end{acknowledgements}

\newcommand{\npb}{Nucl. Phys.}\newcommand{\adv}{Adv.
  Phys.}\newcommand{\epl}{Europhys. Lett.}

\end{document}